\documentclass[compactaffiliation]{Interspeech}
\usepackage{hyperref}
\usepackage{multirow}
\usepackage{amsmath}
\usepackage{graphicx}
\usepackage{tikz}
\usetikzlibrary{shapes, arrows, positioning}
\interspeechcameraready

\title{Leveraging AM and FM Rhythm Spectrograms for Dementia Classification and Assessment}
\author[affiliation={1,2}]{Parismita}{Gogoi}
\author[affiliation={3}]{Vishwanath Pratap}{Singh}
\author[affiliation={4}]{Seema}{Khadirnaikar}
\author[affiliation={5}]{Soma}{Siddhartha}
\author[affiliation={6}]{Sishir}{Kalita}
\author[affiliation={3}]{Jagabandhu}{Mishra}
\author[affiliation={7}]{Md}{Sahidullah}
\author[affiliation={1}]{Priyankoo}{Sarmah}
\author[affiliation={8}]{S. R. M.}{Prasanna}

\affiliation{}{IIT Guwahati}{India}
\affiliation{DUIET}{Dibrugarh University}{India}
\affiliation{}{University of Eastern Finland}{Finland}
\affiliation{}{Independent Researcher}{India}
\affiliation{}{Saryps Labs}{India}
\affiliation{}{Armsoftech.air}{India}
\affiliation{}{TCG CREST}{India}
\affiliation{}{IIIT Dharwad}{India}

\email{
  parismitagogoi@iitg.ac.in, jagabandhu.mishra@uef.fi
}

\keywords{Alzheimer's dementia, Dementia, Rhythm formant, Speech pathology}

\usepackage{comment}

\begin{document}

\maketitle

% the abstract here must exactly match the abstract entered into the paper submission system
\begin{abstract}
This study explores the potential of Rhythm Formant Analysis (RFA) to capture long-term temporal modulations in dementia speech. Specifically, we introduce RFA-derived rhythm spectrograms as novel features for dementia classification and regression tasks. We propose two methodologies: (1) handcrafted features derived from rhythm spectrograms, and (2) a data-driven fusion approach, integrating proposed  RFA-derived rhythm spectrograms with vision transformer (ViT) for acoustic representations along with BERT-based linguistic embeddings. We compare these with existing features. Notably, our handcrafted features outperform eGeMAPs with a relative improvement of $14.2\%$ in classification accuracy and comparable performance in the regression task. The fusion approach also shows improvement, with RFA spectrograms surpassing Mel spectrograms in classification by around a relative improvement of $13.1\%$ and a comparable regression score with the baselines.  All codes are available in GitHub repo\footnote{\url{https://github.com/seemark11/DhiNirnayaAMFM}}.
% We propose a framework for detecting and assessing dementia by analyzing deviations in rhythmic patterns within speech. 
% %We propose a multimodal framework for dementia detection that integrates acoustic and linguistic features from natural speech. 
% Our approach employs rhythm formant analysis (RFA) to extract low-frequency (below 10 Hz) spectrograms from amplitude modulation and frequency modulation envelopes, capturing long-term rhythmic variations. We assess the effectiveness of the proposed RFA features by comparing them with two existing feature-based systems: (1) System-1: … (details about the system) …, and (2) System-2: a state-of-the-art Mel spectrogram-based system designed for classification and regression tasks. Our results show that the proposed RFA features achieve a relative improvement of x\% and y\% in classification performance over System-1 and System-2, respectively. For the regression task, we observe a relative reduction of x\% and y\% in mean squared error compared to System-1 and System-2, respectively.   %These RFA spectrograms are concatenated with log-Mel spectrograms (which capture short-term spectral details) into an RGB image that is fed to a Vision Transformer (ViT) for robust acoustic embedding extraction.    
\end{abstract}

\section{Introduction}
Dementia describes a cluster of neurodegenerative conditions characterized by progressive cognitive decline, with Alzheimer’s disease (AD) being the most prevalent cause~\cite{WHO2023Dementia}. While memory loss is often considered the primary clinical hallmark, speech, and language impairments emerge early and can manifest as hesitations, disrupted rhythm, word-finding difficulties, and prosodic changes~\cite{klimova2016speech}. These early linguistic and paralinguistic markers have motivated various speech-based approaches for dementia assessment, offering a non-invasive, cost-effective alternative to traditional neuroimaging and clinical testing~\cite{luz2021alzheimer,satt2013evaluation}.

%Dementia, as it progresses, introduces artefacts in speech production, affecting various stages—from message generation to reduced control over vocal filter articulators due to motor control impairments. These artifacts, which reflect the severity of dementia, are present across all aspects of speech production. Keeping this in mind, the literature derives these artefacts from speech or spoken text regarding acoustic and linguistic evidence~\cite{}.  
Speech researchers have been investigating spoken language, both acoustically and linguistically, as key evidence for detecting and analyzing dementia~\cite{cho2022lexical}. Handcrafted linguistic features such as part-of-speech patterns, type-token ratio, hesitation-related features, vocabulary variation~\cite{cho2022lexical}, and syntactic complexity~\cite{tomoeda1990speech} have been widely explored. Additionally, data-driven features derived from models such as, BERT~\cite{pan2021using} and multilayer bidirectional transformer encoders~\cite{pan2025two} are also used for dementia assessment. Similarly, acoustic features including speech rate~\cite{saeedi2024acoustic}, fundamental frequency~\cite{cho2022lexical}, rhythm, Mel-frequency cepstral co-coefficients (MFCC), and extended Geneva Minimalistic Acoustic Parameter Set (eGeMAPs)~\cite{chen2021automatic} have been employed. More recently, speech representations extracted from foundation models, e.g., Wav2Vec2.0 and vision transformers (ViT)~\cite{ilias2023detecting} have gained attention for dementia detection. Furthermore, studies suggest that linguistic and acoustic features offer complementary evidence for dementia assessment~\cite{ilias2023detecting}.

\begin{figure*}[t]
\begin{center}
\includegraphics[width=\textwidth]{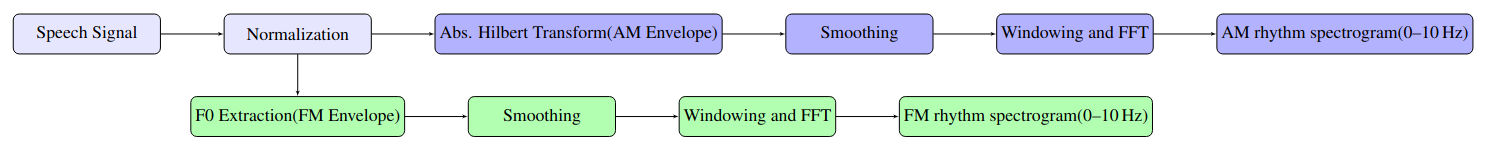}
\caption{Block diagram of the AM (blue) and FM (green) rhythm spectrogram computation 
pipelines.}\label{fig:AM_FM_spectrogram}
\end{center}
\vspace{-0.5cm}
\end{figure*}

Rhythmic analysis of continuous speech---that is, examining how syllables and words are temporally distributed---has shown promise for dementia detection~\cite{oh2021systematic}. Prior studies report that individuals with dementia often exhibit longer pauses, reduced articulation rates, decreased intensity variability, and altered rhythmic structures~\cite{oh2021systematic, oh2022prosody, nevler2017automatic, cho2022lexical}. Consequently, acoustic features such as articulation rate, word rate, syllable rate, and pause rate have been widely investigated as markers of cognitive deterioration~\cite{cho2022lexical}.

In this study, we build upon these fundamental findings by utilizing \emph{rhythm formant analysis} (RFA)~\cite{gibbon2021jipa} to capture long-term temporal modulations embedded in the speech signal. The RFA focuses on the low-frequency (LF) components (below 10~Hz) of both amplitude modulation (AM) and frequency modulation (FM) envelopes, thereby revealing prosodic and articulatory variations that evolve over time~\cite{gibbon2021jipa, gibbon18_speechprosody, Gibbon2019quantify}. Unlike conventional temporal rhythm analysis---which depends on syllable- or word-level annotations (often requiring manual effort or forced alignment)---RFA is entirely annotation-free~\cite{gibbon2021jipa}. This is especially beneficial in pathological speech processing, where accurate manual annotation is challenging and demands specialized linguistic expertise. RFA characterizes rhythm by detecting spectral peaks, known as \emph{rhythm formants}, from the LF spectrum, rather than relying solely on the duration of individual speech units~\cite{gibbon2021jipa, Parismita}. Furthermore, \emph{rhythm spectrogram} has been introduced in RFA to analyze long-term rhythmic patterns by leveraging the temporal details in the AM and FM envelopes of speech utterances~\cite{gibbon2021jipa, parismita_ijalp}. This modulation-theoretic approach provides an inductive method to capture the evolving nature of rhythm. Moreover, RFA provide a dynamic visualization of these long-term rhythmic patterns and capture information that correlates with traditional measures such as syllable and word rates~\cite{Parismita, gibbon2021jipa}. However, the effectiveness of RFA in detecting or assessing disordered speech, such as that of individuals with dementia, remains largely unknown, with no prior attempts reported in the literature.

%Furthermore, the concept of rhythm spectrograms generated via RFA provide a dynamic visualization of these long-term rhythmic patterns and capture information that correlates with traditional measures such as syllable and word rates~\cite{Parismita, gibbon2021jipa}. However, effectiveness of RFA in detecting or assessing disorder speech such as speech of individuals with dementia is largely unknown and no attempt has been made in the literature. 
%Notably, the AM and FM LF spectrograms encode complementary discriminative information, and our analysis indicates that fusing them can enhance the detection of dementia-related speech changes.

Driven by the significance of RFA in analyzing long-term rhythm, we hypothesize that RFA-derived spectrograms can capture rhythmic deviations of speech in individuals with dementia. Moreover, it is observed that the audio from the elicitation tasks, such as the \emph{Cookie Theft Picture Description} task, as of several seconds duration. The long duration spontaneous speech file contains the changes in rhythm/prosody of the speaker over the time and it's the idea for RFA-based frequency domain rhythm analysis. This work presents two methodology for detecting and assessing dementia based on the characterization of AM and FM rhythm spectrograms. In the first approach, we compute the variance of rhythm formants over time from the rhythm spectrograms and explore the 2D discrete cosine transform-based joint spectro-temporal representation of rhythm spectrograms. These handcrafted features are then fed into a machine learning classifier to detect dementia and predict the corresponding Mini-Mental Status Examination (MMSE) score of the speaker. In the second approach, we investigate a vision transformer (ViT)-based data-driven acoustic representation of rhythm spectrograms and integrate it with a BERT-based linguistic representation to enhance dementia detection and MMSE score estimation.

\section{Rhythm spectrogram computation}
\begin{figure}[t]
\begin{center}
\includegraphics[width=3in]{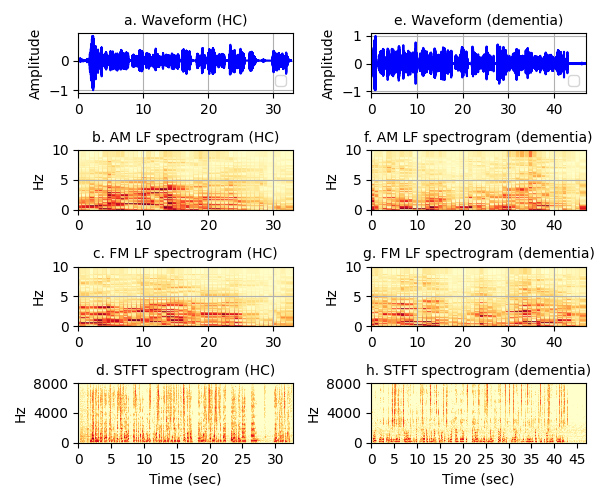}
\caption{Illustration of AM and FM rhythm spectrograms of speech utterance from healthy control (HC) and dementia.}\label{fig:fmam}
\end{center}
\vspace{-0.8cm}
\end{figure}
We use the RFA method reported in~\cite{gibbon2021jipa} to compute the AM and FM rhythm spectrograms. The overall block diagram illustrating the computation process is shown in Figure~\ref{fig:AM_FM_spectrogram} and described as follows.

\textbf{Computation of AM rhythm spectrogram.}
The speech signal is first normalized using its maximum absolute value. An absolute Hilbert transform is applied to obtain the AM envelope. The resulting AM envelope is smoothed to reduce rapid fluctuations. A $5$ s window with overlapping steps is used to extract a fixed set of $100$ segments from the AM envelope, ensuring consistent temporal segmentation regardless of variations in utterance duration. Each segment is transformed via FFT to capture low-frequency components of the envelope. The short-term FFT magnitude spectra for each $5$\,s window are stacked in time order to form a time-frequency representation (i.e., the AM rhythm spectrogram). Only frequencies in the $0$--$10$\,Hz range are retained, excluding the DC component ($0$\,Hz), and the spectral amplitudes are normalized. 

\textbf{Computation of FM rhythm spectrograms.}
The fundamental frequency ($F_{0}$) contour of the speech signal is computed using the RAPT pitch-tracking algorithm \cite{talkin1995robust} (via the \texttt{pysptk} \cite{yamamoto2019r9y9} Python package). The $F_{0}$ contour is smoothed to produce the FM envelope. Voiceless segments and pauses appear as breaks in the contour but are preserved as valid components for subsequent spectral analysis. As with the AM envelope, a $5$ s window with overlapping steps is used to extract a fixed set of $100$ segments from the FM envelope. The resulting spectra from each segment are concatenated over time to yield the FM rhythm spectrogram within the $0$--$10$\,Hz range. The DC component is discarded, and amplitudes are normalized.

Figure~\ref{fig:fmam} illustrates the AM and FM rhythm spectrograms for speech utterances from healthy controls (HC) and individuals with dementia. The rhythmic patterns differ clearly between the two groups. These differences suggest that analyzing rhythm spectrograms may provide valuable cues for detecting dementia.

\section{Proposed approach for dementia detection and assessment}
We employ AM and FM rhythm spectrograms for dementia detection and assessment using two approaches: (1) extracting handcrafted features for classification and regression with machine learning models, and (2) leveraging a data-driven approach with the ViT-BERT acoustic-linguistic end-to-end (E2E) fusion model~\cite{ilias2023detecting}. The embeddings extracted from ViT-BERT are further used for regression with machine learning models. For classification, we use a support vector machine (SVM) classifier, while regression is performed using SVM and decision tree (DT) regression.  These methods are selected based on prior studies demonstrating the superior performance of SVM for classification~\cite{luz2021detecting} and both SVM and DT for regression~\cite{luz2021detecting} in dementia detection and assessment.

\subsection{Handcrafted characterization of rhythm spectrograms for classification and regression}
In this work, we extract the $N$ rhythm formants from each LF spectrum slice of the spectrogram using the peak-picking algorithm described in~\cite{2020SciPy}. These rhythm formants are then tracked over time, producing trajectories that capture changes in rhythm. The variance of these rhythm formant trajectories provides an interpretable measure of rhythmic variation. Therefore, for each utterance, there are $2N$ variance-based rhythm values ($N$ for AM-based spectrogram and $N$ for FM-based spectrogram).

Along with this, we also compute the two-dimensional discrete cosine transform (2D-DCT)~\cite{rao1990dct, SofiaGoodnessMisarticulated2015, kalita2018intelligibility, parismita_ijalp} of the AM and FM rhythm spectrograms to capture spectro-temporal variations directly from their time-frequency representations. After computing 2D-DCT, we consider only the lower-order coefficients by selecting the first $C$ vertical and horizontal DCT coefficients, forming a $C \times C$ matrix. Flattening this matrix yields a $C^2$-dimensional feature vector. Considering both AM and FM rhythm spectrograms, we obtain a total of $2\times C^2$ DCT features.

By combining both variance and 2D-DCT-based features, each utterance is represented by a $2N + 2\times C^2$-dimensional feature vector, which is subsequently used for classification and regression with machine learning models.

%A detailed discussion on the importance of 2D-DCT in join spectro-temporal represenrations from spectrogram and equations for it's computation can be found in~\cite{kalita2018intelligibility,SofiaGoodnessMisarticulated2015,parismita_ijalp}. In this work, to obtain a compact representation, we have considered only the lower-order coefficients by selecting the first two vertical and two horizontal DCT coefficients, forming a \(2 \times 2\) matrix. Flattening this matrix yields a 4-dimensional feature vector. Therefore, combining for AM and FM LF spectrograms, we get 8 dimensional 2D-DCT feature vector.

\subsection{ Data-driven characterization of rhythm spectrogram using ViT-BERT for classification}

The ViT-BERT fusion model, trained using both acoustic and linguistic evidence, has recently been used in ~\cite{ilias2023detecting} for dementia classification, demonstrating superior performance compared to using only acoustic or linguistic evidence. Inspired by this study, we replace the mel-spectrogram and its $\Delta$, $\Delta\Delta$ with AM and FM rhythm spectrograms and their $\Delta$, $\Delta\Delta$ to highlight the relevance of rhythm-based features. Furthermore, instead of relying on ground truth transcripts, we use transcripts generated by automatic speech recognition (ASR) to extract linguistic information. This modification accounts for practical scenarios where obtaining manual transcriptions during testing is challenging~\cite{chellenging-transcriptions}. 

\begin{figure}[t]
\centering
\includegraphics[width=2.7 in, height=3.8 in]{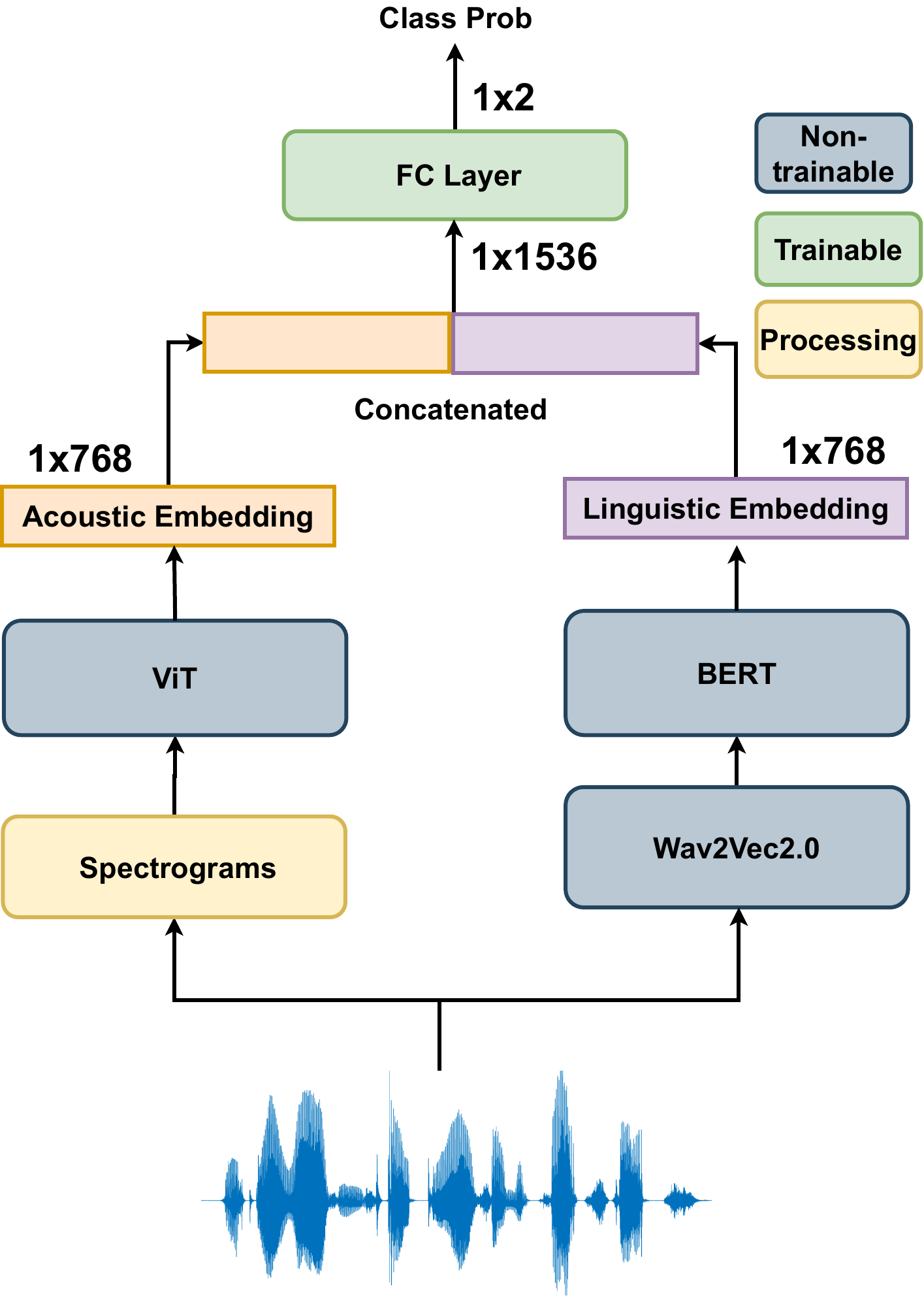}
\caption{End-to-end pipeline for dementia detection using ViT-BERT fusion system.}
\vspace{-0.6 cm}
\label{fig:sys-2}
\end{figure}

%The block diagram of the ViT-BERT system is depicted in Figure~\ref{fig:sys-2}. Given a speech utterance, the AM and FM rhythm spectrograms are computed on the acoustic-feature related branch, while the linguistic-feature related branch utilizes widely used self-supervised Wav2Vec2.0 ASR to transcribe the audio~\cite{baevski2020wav2vec}. The AM and FM spectrograms, along with their $\Delta$ and $\Delta\Delta$ spectrograms, are each provided as three input channels to ViT. Similarly, the transcripts generated by ASR serve as input to BERT. We employ pre-trained, open-source ViTS~\footnote{\url{https://huggingface.co/timm/vit_base_patch16_224.augreg_in21k}}, Wav2Vec2.0~\footnote{\url{https://huggingface.co/facebook/wav2vec2-large-960h}}, and BERT~\footnote{\url{https://huggingface.co/google-bert/bert-base-uncased}} models as non-trainable components in our architecture. The outputs of ViT and BERT are concatenated and fed into a trainable fully connected layer with two neurons. The total number of trainable parameters in our architecture is $3072$.

The block diagram of the ViT-BERT system is depicted in Figure~\ref{fig:sys-2}. Given a speech utterance, the AM and FM rhythm spectrograms are computed on the acoustic-feature-related branch, while the linguistic-feature-related branch utilizes the widely used self-supervised Wav2Vec2.0 ASR to transcribe the audio~\cite{baevski2020wav2vec}. Wav2Vec2.0 is a transformer-based model trained on large-scale speech data in a self-supervised manner for robust transcription. The AM and FM spectrograms, along with their $\Delta$ and $\Delta\Delta$ spectrograms, are each provided as three input channels to ViT. ViT excels at capturing spatial and temporal dependencies in image-like data, making it particularly suited for processing spectrogram representations of speech. Similarly, the transcripts generated by ASR serve as input to BERT, a transformer-based language model that captures contextual relationships within text, enabling a richer understanding of linguistic structure and meaning.

We employ pre-trained, open-source ViT~\footnote{\url{https://huggingface.co/timm/vit_base_patch16_224.augreg_in21k}}, Wav2Vec2.0~\footnote{\url{https://huggingface.co/facebook/wav2vec2-large-960h}}, and BERT~\footnote{\url{https://huggingface.co/google-bert/bert-base-uncased}}  models as non-trainable components in our architecture. The outputs of ViT and BERT (each 768 dimensions) are concatenated and fed into a trainable fully connected layer with two neurons, facilitating joint learning from acoustic and linguistic modalities. The total number of trainable parameters in our architecture is $3074$.

% Please add the following required packages to your document preamble:
% \usepackage{multirow}
% \usepackage{graphicx}

%==========================================================
\section{Experimental setup and results}

\subsection{Dataset description}
We used Alzheimer’s Dementia Recognition through Spontaneous Speech Only (ADReSSo)~\cite{luz2021detecting} in this study. This dataset was originally introduced as part of the ADReSSo Challenge 2021~\cite{luz2021detecting}, aiming to detect dementia and assess cognitive decline using speech alone. It consists of audio recordings of participants, both with and without Alzheimer’s dementia, describing the “Cookie Theft” scene from the Boston Diagnostic Aphasia Examination, as well as recordings from a verbal fluency task. The dataset was carefully balanced by age and gender to minimize confounding variables. The dataset contains a total of $237$ audio recordings, with $166$ used for training and $71$ for testing. In addition to speech utterances, the dataset provides transcriptions and timestamps of conversations between subjects and interviewers. In our study, we use these timestamps to isolate only the subject’s speech segments, concatenate them for each utterance, and use the resulting utterance for further processing.

\subsection{Experimental setup}
Apart from our proposed approaches, we consider the eGeMAPS~\cite{eyben2015geneva} features to compare the performance of our handcrafted acoustic features in dementia detection using an SVM classifier. Since the ADReSSo dataset does not provide a predefined development set, we perform all experiments using a $5$ fold split of the ADReSSo training set into training and development subsets, maintaining an $80:20$ ratio. For machine learning models used in classification and regression, the optimized hyperparameters are determined through cross-validation, and the training parameters are averaged across the $5$ folds. The resulting averaged model is then used for inference.

\textbf{Handcrafted characterization of rhythm spectrogram.}
The rhythm variance feature is computed using $N=6$ formants, while the 2D-DCT coefficients are extracted with $C$ varying from $2$ to $4$ for classification and regression. These features are then used with an SVM classifier for classification and with SVR and DT for regression. 

\textbf{ViT-BERT system using rhythm spectrogram.}
Following the system proposed in~\cite{ilias2023detecting}, we train three ViT-BERT models by varying the input acoustic features. The first system, which uses a Mel-spectrogram as reported in~\cite{ilias2023detecting}, serves as the baseline for our study. The other two systems use the AM rhythm spectrogram and FM rhythm spectrogram, respectively. The models are trained using cross-entropy loss with a $5$ fold cross-validation. The trained parameters are averaged, and used for inference. Furthermore, to ensure reliability, training is repeated with three fixed random seeds, and we report the mean and standard deviation of the results. The trained model is used directly for classification, whereas the concatenated embeddings extracted from the trained model are used to train the DT and SVR regression models for the regression task to predict the MMSE score.

% \subsection{Baseline feature representation}
% % For citation see this paper: https://www.pure.ed.ac.uk/ws/files/122192019/JSTSP.pdf
% For our baseline acoustic representation, we use the eGeMAPS [35] feature set, which was developed to offer a concise set of descriptors that capture physiological voice characteristics shown to be predictive of cognitive changes [60]. eGeMAPS covers F0 semitone, loudness, spectral flux, MFCC, jitter, shimmer, formants (F1, F2, F3), alpha ratio, Hammarberg index, slope V0, and their common functionals, for a total of 88 features. In our work, we extract these features from the entire utterance as a single vector.

\subsection{Results and discussion}
\subsubsection{Classification systems}
We evaluated the classification system using both handcrafted features with the SVM and ViT-BERT E2E model. The obtained performances are discussed as follows.

\begin{table}[t]
\centering
\caption{Classification results in terms of accuracy (\%) and F1-score (\%) for the SVM-based model.}
\begin{tabular}{|c|c|c|c|}
\hline
 & Variance & 2D-DCT  & Combined \\ \hline
Accuracy               & 62.86    & 65.71 & 65.71    \\ \hline
F1-score               & 68.29     & 64.71  & 69.23     \\ \hline
\end{tabular}
\label{tab:classification_SVC_results}
\end{table}

\textbf{Handcrafted features.} Using handcrafted features, we evaluate our trained SVM model, and the test set results are presented in Table~\ref{tab:classification_SVC_results}. We consider three training and evaluation conditions: (1) using only the variance of rhythm formants, (2) using only 2D-DCT coefficients, and (3) using a combination of both. The variance-based feature alone achieves an accuracy of $62.86\%$ and an F1-score of $68.29\%$. By varying the number of 2D-DCT coefficients ($C$) from $2$ to $4$ and evaluating performance, we observe that $C=3$ yields the best results, with an accuracy of $65.71\%$ and an F1-score of $64.71\%$. The combined feature set achieves the same accuracy ($65.71\%$) as the 2D-DCT features but improves the F1-score to $69.23\%$. Interestingly, while 2D-DCT features yield better accuracy, the variance-based features provide a higher F1-score. The combined system maintains the accuracy of 2D-DCT while further improving the F1-score.

\begin{table}[t]
\centering
\caption{Results ViT-BERT, with Mel, and AM and FM rhythm spectrogram and their $\Delta$ and $\Delta\Delta$ inputs to ViT, C1, C2, C3 indicate $3$ channels of ViT input. Results are in the form of mean $±$ standard deviation over $3$ different runs.}
\label{tab:sys2}
\resizebox{\columnwidth}{!}{%
\begin{tabular}{|c|ccc|cc|}
\hline
\multirow{2}{*}{} & \multicolumn{3}{c|}{Input Features to ViT}            & \multicolumn{2}{c|}{Results}       \\ \cline{2-6} 
                  & \multicolumn{1}{c|}{C1} & \multicolumn{1}{c|}{C2} & C3 & \multicolumn{1}{c|}{Accuracy} & F1 \\ \hline
\textbf{Baseline~\cite{ilias2023detecting}} &
  \multicolumn{1}{c|}{Mel} &
  \multicolumn{1}{c|}{$\Delta$} &
  $\Delta\Delta$ &
  \multicolumn{1}{c|}{73.33 ± 0.67} &
  {72.98} ± 0.006 \\ \hline
\multirow{2}{*}{\textbf{Proposed}} &
  \multicolumn{1}{c|}{FM} &
  \multicolumn{1}{c|}{$\Delta$} &
  $\Delta\Delta$ &
  \multicolumn{1}{c|}{71.43 ± {0.10}} &
  71.10 ± 0.038 \\ \cline{2-6} 
 &
  \multicolumn{1}{c|}{AM} &
  \multicolumn{1}{c|}{$\Delta$} &
  $\Delta\Delta$ &
  \multicolumn{1}{c|}{\textbf{74.29} ± {0.23}} &
  \textbf{74.06} ± {0.002} \\ \hline
\end{tabular}%
}
\end{table}

\textbf{ViT-BERT System.} We evaluated three systems (1) Mel spectrogram, (2) AM, and (3) FM rhythm spectrogram along with their $ \Delta$ and $\Delta\Delta$ as acoustic representation fed into the three channels of the ViT model. Our baseline—using Mel, Delta Mel, and double Delta Mel features—achieved a mean accuracy of $73.33\%$ and an F1-score of $72.98\%$ over three runs, as reported in~\cite{ilias2023detecting}. In comparison, incorporating FM, Delta FM, and double Delta FM features yielded $71.43\%$ accuracy and a mean F1-score of $71.10\%$. Notably, the best performance was obtained when using AM, Delta AM, and double Delta AM features, which achieved $74.29\%$ accuracy and an F1-score of $74.06\%$. In all experiments, the input to BERT consisted of Wav2Vec2.0-based ASR transcriptions. Overall, the proposed AM features provide a relative $13.09\%$  improvement in accuracy over the baseline.

Finally, the performance comparison of both handcrafted and data-driven characterizations of rhythm spectrograms against their respective baselines—$88$-dimensional eGeMAPS for handcrafted features and Mel-spectrogram-based ViT-BERT for data-driven features—is presented in Table~\ref{tab:com-class}. The results demonstrate that rhythm spectrograms consistently outperform their respective baselines in both approaches, highlighting their effectiveness in capturing dementia-related speech patterns. These findings underscore the potential of rhythm spectrograms in encapsulating valuable evidence for dementia detection.

% Please add the following required packages to your document preamble:
% \usepackage{graphicx}
\begin{table}[t]
\centering
\caption{Comparison of classification results obtained from the baseline and proposed systems.}
\label{tab:com-class}
\resizebox{\columnwidth}{!}{%
\begin{tabular}{|c|cc|cc|}
\hline
         & \multicolumn{2}{c|}{Handcrafted features}   & \multicolumn{2}{c|}{Data-driven features}   \\ \hline
 & \multicolumn{1}{c|}{Variance+2D-DCT} & eGeMAPS ~\cite{luz2021detecting} & \multicolumn{1}{c|}{AM spectrogram} & Mel spectrogram \\ \hline
Accuracy (\%) & \multicolumn{1}{c|}{\textbf{65.71}} & 64.79 & \multicolumn{1}{c|}{\textbf{74.29}} & 73.33 \\ \hline
F1-score (\%) & \multicolumn{1}{c|}{\textbf{69.23}}  & -     & \multicolumn{1}{c|}{\textbf{74.06}}  & 72.98  \\ \hline
\end{tabular}%
}
\end{table}

\subsubsection{Regression systems}
Table~\ref{tab:reg_results} reports root mean squared error (RMSE) and Pearson correlation coefficient (\(\rho\)) for MMSE estimation using regression models trained on handcrafted features (variance and 2D-DCT) and ViT-BERT-based embeddings.

\begin{table}[]
\centering
\caption{Regression results for MMSE estimation in terms of root mean squared error (RMSE) and Pearson correlation coefficient (\(\rho\)), SVR: support vector regression, DT: decision tree.}  
\resizebox{\columnwidth}{!}{
\begin{tabular}{|c|c|c|c|c|}
\hline
Model    & Variance    & 2D-DCT        & Combined    & Embeddings  \\ \hline
SVR & 6.50 (0.23) & 6.39 (0.29) & 6.26 (0.37) & \textbf{6.00 (0.42)} \\ \hline
DT  & 7.34 (0.25) & 8.84 (0.01)    & 7.64 (0.32) & 6.57 (0.27) \\ \hline
\end{tabular}}
\label{tab:reg_results}
\end{table}

\textbf{Handcrafted features.} Among the handcrafted features, SVR yields the lowest RMSE of $6.39$ when the lower-order coefficient matrix of the 2D-DCT coefficient $C$ is set to $2$. Consistent with the classification results, the 2D-DCT feature outperforms the variance-based features in the regression task. 
Furthermore, combining the variance and 2D-DCT features improves performance compared to using either feature individually (RMSE: $6.26$, \(\rho\): $0.37$). However, the eGeMAPS features achieve a lower RMSE of $6.09$, as reported in~\cite{luz2021detecting}.

\textbf{ViT-BERT embeddings.} For the ViT-BERT embeddings, we consider only those derived from the AM rhythm spectrogram, as it demonstrated superior performance in the classification task. As shown in Table~\ref{tab:reg_results}, SVR outperforms DT, achieving an RMSE of $6.00$ with \(\rho\) = $0.42$, compared to the DT model’s RMSE of $6.57$ with \(\rho\) = $0.27$. These results suggest that embedding-based features perform comparably to, or slightly better than, eGeMAPS features for MMSE score estimation. 

%==========================================================
\section{Conclusions}
In this study, we explored the use of rhythm spectrograms through both handcrafted and data-driven representations for dementia detection. Experimental results demonstrate that the proposed characterization of rhythm spectrograms achieves superior performance in dementia classification and comparable results in MMSE score prediction. In the future, we plan to integrate these features with existing approaches to evaluate their combined effectiveness in distinguishing between dementia and non-dementia cases. Additionally, we aim to benchmark the proposed features across multiple datasets to further validate their robustness and generalizability.

%These findings highlight the signifacant role of rhythm spectrograms in dementia detection. The AM envelope effectively encodes the sonority contours of speech—capturing information about syllables, words, phrases, and long-term features such as pauses—while the FM envelope represents intonation and tonal variations~\cite{gibbonComp, gibbon2021jipa}. In Alzheimer’s disease speech, pronounced deviations in these patterns lead to enhanced discrimination when using rhythm-based features.

%\section{Acknowledgments}

\section{Acknowledgements}
The authors wish to acknowledge CSC – IT Center for Science, Finland, for computational resources.

\bibliographystyle{IEEEtran}
\bibliography{mybib}

\end{document}